\documentclass[aps,floatfix,nofootinbib,prd]{revtex4-1}
 \usepackage{graphicx} 

\newcommand{\eq}[1]{(\ref{#1})}
\newcommand{\be}{\begin{equation}}
\newcommand{\ee}{\end{equation}}
\newcommand{\bea}{\begin{eqnarray}}
\newcommand{\eea}{\end{eqnarray}}

\newcommand{\vs}[1]{\vspace{#1 mm}}
\newcommand{\hs}[1]{\hspace{#1 mm}}

\def\cc{\gamma}

\def\d{\delta}

\def\e{\epsilon}

\def\f{\phi}
\def\fr{\frac}
\def\F{\Phi}
\def\vf{\varphi}

\def\l{\lambda}

\def\m{\mu}
\def\n{\nu}

\def\s{\sigma}

\def\t{\tau}
\def\th{\theta}

\def\z{\zeta}

\def\del{\partial}

\let\bm=\bibitem
\def\nn{\nonumber}

\begin{document}
\large

\title{A (semi)-exact Hamiltonian for the curvature perturbation $\zeta$}

\author{Ali Kaya}

\email[]{alikaya@tamu.edu}
\affiliation{\vs{3}Department of Physics and Astronomy, Texas A\&M University, College Station, TX 77843, USA}

\begin{abstract}

The total Hamiltonian in general relativity, which involves the first class Hamiltonian and momentum constraints, weakly vanishes. However, when the action is expanded around a classical solution as in the case of a single scalar field inflationary model, there appears a non-vanishing Hamiltonian and additional first class constraints; but this time the theory becomes perturbative in the number of fluctuation fields. We show that one can reorganize this expansion and solve the Hamiltonian constraint exactly, which yield an explicit all order action. On the other hand, the momentum constraint can be solved perturbatively in the tensor modes $\gamma_{ij}$ by still keeping the curvature perturbation $\zeta$ dependence exact. In this way, after gauge fixing, one can obtain a semi-exact Hamiltonian for $\zeta$ which only gets corrections from the interactions with the tensor modes (hence the Hamiltonian becomes exact when the tensor perturbations set to zero). The equations of motion clearly exhibit when the evolution of $\zeta$ involves a logarithmic  time dependence, which is a subtle point that has been debated in the literature. We discuss the long wavelength and late time limits, and obtain some simple but non-trivial classical solutions of the $\z$ zero-mode. 

\end{abstract}

\maketitle

\section{Introduction}

Determining the evolution of the cosmological perturbations starting from the moment they appeared as sub-horizon quantum fluctuations to the much later epochs involving structure formation is crucial in testing the predictions of inflationary models. Although the problem is basically an expansion around a classical background solution, there are important technical issues to be tackled like the gauge invariance related to the coordinate transformations. The cosmological perturbation theory (see the classical review \cite{mfb}) properly classifies the fluctuations and determines their basic dynamical evolution like the freeze-out of superhorizon modes. In recent years, after the seminal work of \cite{mal}, there has been a great interest in determining the evolution of the cosmological perturbations beyond the linear regime, which has important implications for the precision cosmology like the presence of non-gaussianities. In the approach of \cite{mal}, one works in the Lagrangian formulation and takes the metric in the Arnowitt-Deser-Misner (ADM) form. One then algebraically solves for the lapse $N$ and the shift $N^i$ from their equations of motion and imposes gauge fixing to obtain a Lagrangian for the physical degrees of freedom, i.e the curvature perturbation $\z$ and the (transverse, trace-free) tensor mode $\cc_{ij}$ describing the gravitational waves. This procedure, which can be utilized perturbatively, becomes quite cumbersome at higher orders. Yet, one can systematically work out the quantum corrections to cosmological correlation functions (including the loop effects) by using the in-in (Schwinger-Keldysh) formalism \cite{w1,w2}. 

Although perturbation theory seems to work well in dealing with small non-linear effects, it is also desirable to have non-perturbative/exact methods to understand the implications of possible strong non-linearities.  One can, for example, utilize the symmetries of the background geometry to obtain exact Ward identities for cosmological correlation functions \cite{ward1,ward2}, see also \cite{ward-ali}. Here, the dilation symmetry gives the well-known three-point consistency condition \cite{mal,3pt}. The stochastic approach of \cite{np1,np2}  can be used to calculate arbitrary large correlation functions in de Sitter space, which involve the so called infrared (IR) logarithms (see also \cite{ir1,ir2}, and also \cite{ak2} which shows the existence of IR logarithms in the minisuperspace approximation). It is also possible to utilize the exact renormalization group flow techniques for quantum fields to infer the structure of the effective interacting potential of a real scalar field in de Sitter space \cite{akr}. 

More recently, using powerful physical principles like locality, unitarity and symmetry,  the cosmological bootstrap strategy has been used to constrain correlation functions, for review see e.g. \cite{boot1,boot2}. This program offers a very broad, model independent picture of inflationary physics and allows one to  infer various exact results and new insights. There are numerous important findings that follow from the bootstrap program like proving a cosmological optical theorem, which implies  an infinite set of relations among the correlators \cite{ot1},  and obtaining cosmological cutting rules that fix the discontinuity of a loop  diagram in terms of lower order loops and tree diagrams \cite{ot2}. Another interesting line of research involves fixing the correlators of massless spin 1 and spin 2 fields  in de Sitter space by solving the Ward identities related to background symmetries \cite{nr1}. 

There are also proposals for a possible holographic dual  description of single scalar field inflationary models,  where the dual is a three dimensional quantum field theory (a conformal field theory deformed by a nearly marginal operator) that flows from an IR fixed point to a UV fixed point corresponding in the dual picture to the (hilltop) inflaton rolling from a local maximum at past infinity to a local minimum at future infinity \cite{hi1,hi2}. 

Compared to these non-perturbative methods, our approach in this paper is more direct, i.e.  we only consider  single scalar field inflationary models and work in the context of classical general relativity. As mentioned above, the action for the cosmological perturbations arises from an expansion around a background solution and it contains infinitely many interaction terms that can in principle be determined at any desired order by a standard but lengthy procedure. Moreover, the expansion is naturally carried out in the Lagrangian formulation and one should obtain the corresponding Hamiltonian for the in-in perturbation theory, which is an additional involved computation. Obviously, it is desirable to have a closed exact expression for the Hamiltonian of the physical fluctuation variables. In this paper we show that one can indeed obtain a non-perturbative Hamiltonian for the curvature perturbation $\z$ by paying attention how the expansion around the background cosmological solution yields a non-zero fluctuation Hamiltonian.  As we will see both the special structure of the constraints in general relativity and the particular form of the cosmological background become important in this all-order derivation, which in the first step amounts to solving the Hamiltonian constraint exactly. The notorious problem of time that appears when dealing with the Hamiltonian constraint is eliminated by referring to the background solution. To fix the dynamics of the  physical degrees of freedom, one must also solve  the momentum constraint and we show that this can still be done exactly for $\z$ and perturbatively in $\cc_{ij}$. After gauge fixing, the procedure yields a semi-exact Hamiltonian for $\z$ that only gets corrections from $\z$-$\cc_{ij}$ interactions and hence becomes exact when one truncates the system by setting $\cc_{ij}$ (and the corresponding momentum field) to zero.  

As in the original derivation given in \cite{mal}, here also the auxiliary variables coming from the momentum constraint contain non-local expressions involving the Green function of the spatial Laplacian. In this work, we will elaborate on the importance of boundary conditions in obtaining a long wavelength limit or a derivative expansion. As we will see, in the naive long wavelength limit one should treat the momentum constraint exactly since all the auxiliary terms it generates in the action have derivative dimension zero. Nevertheless, we show that the zeroth order derivative expansion can be related to the minisuperspace approximation when suitable boundary conditions are imposed. We determine the late time form of the exact Hamiltonian and obtain simple non-trivial classical solutions of the $\z$ zero-mode corresponding to appropriate initial conditions. Our exact treatment reveals how the so called infrared logarithms emerge in the evolution of $\z$, which has been a subtle point and a source of discussion in the literature, see e.g. \cite{zz2,zz1}. We observe that the exact $\z$ Hamiltonian contains a linear term that explicitly generates $\ln(a)$ dependence for $\z$, however this term is canceled out when one makes a perturbative expansion in the number of fields, hence it is only effective non-perturbatively. On the other hand,  the nonlinear evolution of $\z$ in the minisuperspace model involves highly nontrivial dynamics depending on the initial conditions and  $\ln(a)$ type of time dependence in some cases. 

\section{Main idea}

We consider general relativity in the presence of a minimally coupled self-interacting real scalar field $\f$ that has the potential $V(\f)$. When the metric is taken in the ADM form 
\be\label{13}
ds^2=-N^2dt^2+h_{ij}(dx^i+N^i dt)(dx^j+N^j dt),
\ee
the Einstein-Hilbert action can be written as (we set $8\pi G=1$) 
\be\label{eh1}
S=\int dt\, d^3x\left[\Pi^{ij}\dot{h}_{ij}+P_\f\dot{\f}-N\F-N^i\F_i\right],
\ee
\bea
&&\F=\fr{2}{\sqrt{h}}\left[\Pi^{ij}\Pi_{ij}-\fr12\Pi^2\right]+\fr{1}{2\sqrt{h}}P_\f^2+\sqrt{h}\left[V(\f)+\fr12 h^{ij}\del_i\f\del_j\f-\fr12 R^{(3)}\right],\nn\\
&&\F_i=-2\sqrt{h}D_j\left[\fr{1}{\sqrt{h}}\Pi^{j}{}_i\right]+P_\f\del_i\f,\label{c1}
\eea
where the dot denotes time derivative, $D_i$ is the covariant derivative and $R^{(3)}$ is the Ricci scalar of $h_{ij}$, $h=\det(h_{ij})$ and  $\Pi=\Pi^{ij}h_{ij}$. The spatial indices (like $i$ and $j$) are manipulated by $h_{ij}$. In the classical theory, the canonical pairs obey the Poisson brackets
\bea
&&\left\{h_{ij}(\vec{x}),\Pi^{rs}(\vec{y})\right\}=\fr12\left(\d^r_i\d^s_j+\d^r_j\d^s_i\right)\d^3(\vec{x}-\vec{y}),\nn\\
&&\left\{\f(\vec{x}),P_\f(\vec{y})\right\}=\d^3(\vec{x}-\vec{y}).\label{16}
\eea
The lapse $N$ and the shift $N^i$ are Lagrange multipliers enforcing the Hamiltonian and the momentum constraints; $\F=0$ and $\F_i=0$. 

We introduce the following decomposition of the spatial metric variable
\be
h_{ij}=e^{2\hat{\z}}\,\hat{\cc}_{ij},
\ee
where $\det \hat{\cc}_{ij}=1$ (and thus $h=e^{6\hat{\z}}$). Similarly, the conjugate momenta can be decomposed into the trace and the trace-free parts as   
\be
\Pi^{ij}=\fr16 h^{ij}\,\hat{\pi}+e^{-2\hat{\z}}\,\hat{\s}^{ij},
\ee
where $\hat{\s}^{ij}h_{ij}=0$. In these new variables the action becomes
\be\label{eh2}
S=\int dt\, d^3x\left[\hat{\pi}\dot{\hat{\z}}+\hat{\s}^{ij}\dot{\hat{\cc}}_{ij}+P_\f\dot{\f}-N\F-N^i\F_i\right],
\ee
Note that $(\hat{\z},\hat{\pi})$ and $(\hat{\cc}_{ij},\hat{\s}^{kl})$ are conjugate variables, and one can as usual denote $N^\m=(N,N^i)$ and $\F_\m=(\F,\F_i)$. 

Assume now that there is a cosmological Friedmann-Robertson-Walker (FRW) type solution given by the scale factor $a(t)$ and the scalar field $\f(t)$, and furthermore $N=1$ and $N^i=0$. The background equations of motion 
\bea
&&3\fr{\dot{a}^2}{a^2}=\fr12 \dot{\f}^2+V,\nn\\
&&\fr{\ddot{a}}{a}=-\fr13 \dot{\f}^2+\fr13 V,\label{backsol}\\
&&\ddot{\f}+3\fr{\dot{a}}{a}\dot{\f}+\fr{\del V}{\del \f}=0,\nn
\eea
are thus assumed to hold. The cosmological perturbations can be introduced as fluctuations over the classical background
\bea
&&\hat{\z}=\ln a(t)+\z,\nn\\
&&\f=\f(t)+\vf,\nn\\
&&\hat{\cc}_{ij}=\d_{ij}+\cc_{ij},\nn\\
&&\hat{\pi}=-6a^2\dot{a}+\pi,\nn\\
&&P_\f=a^3\dot{\f}+p_\vf,\label{pert}\\
&&\hat{\s}^{ij}=\s^{ij},\nn\\
&&N=1+n,\nn\\
&&N^i=n^i.\nn	
\eea
The main variables are given by  $(\z,\vf,\cc_{ij},\pi,p_\vf,\s^{ij})$ and $(n,n^i)$ are Lagrange multipliers (one must recall that $\det(\hat{\cc})=\det(\d_{ij}+\cc_{ij})=1$ and $\hat{\cc}_{ij}\s^{ij}=0$). From \eq{16} one can determine the Poisson brackets of these variables as
\bea
&&\{\z(\vec{x}),\pi(\vec{y})\}=\d^3(\vec{x}-\vec{y}),\nn\\
&&\{\cc_{ij}(\vec{x}),\s^{kl}(\vec{y})\}=\fr12\left(\d^k_i\d^l_j+\d^l_i\d^k_j\right)\d^3(\vec{x}-\vec{y})-\fr13\hat{\cc}_{ij}(\vec{x})\hat{\cc}^{kl}(\vec{y})\d^3(\vec{x}-\vec{y}),\label{pb1}\\
&&\{\s^{ij}(\vec{x}),\s^{kl}(\vec{y})\}=\fr13 \hat{\cc}^{kl}(\vec{x})\s^{ij}(\vec{y})\d^3(\vec{x}-\vec{y})-\fr13\hat{\cc}^{ij}(\vec{x})\s^{kl}(\vec{y})\d^3(\vec{x}-\vec{y}),\nn
\eea
where $\hat{\cc}^{ij}$ is the inverse of $\hat{\cc}_{ij}$, and all other brackets vanish; in particular one has
\be
\{\cc_{ij}(\vec{x}),\pi(\vec{y})\}=0, \hs{3}\{\z(\vec{x}),\s^{ij}(\vec{y})\}=0, \hs{3} \{\pi(\vec{x}),\s^{ij}(\vec{y})\}=0.\label{pb2} 
\ee
Using the  expansion \eq{pert} in the action, one obtains 
\be\label{eh3}
S=\int dt\, d^3x\left[\pi\dot{\z}+\s^{ij}\dot{\cc}_{ij}+p_\vf\dot{\vf}-\F^{(\geq2)}-n\left(\F^{(1)}+\F^{(\geq2)}\right)-n^i\F_i^{(\geq1)}\right],
\ee
where we define
\bea
&&\F=\F^{(1)}+\F^{(\geq2)},\nn\\
&&\F_i=\F_i^{(\geq1)}.
\eea
Here, we group the terms in the constraints according to their order;  $\F^{(1)}$ is linear in the perturbations and $\F^{(\geq2)}$ has quadratic and higher order terms. Similarly, $\F_i^{(\geq1)}$ contains linear and higher order terms. Note that the zeroth order values of the constraints, which can be denoted by $\F^{(0)}$ and $\F_i^{(0)}$, vanish by the background equations of motion. By the same reason, the linear fluctuation terms in \eq{eh3} also cancel each other.  

The action \eq{eh3} governing the dynamics of the perturbations $(\z,\vf,\cc_{ij},\pi,p_\vf,\s^{ij})$ has the Hamiltonian (density)\footnote{With some abuse of language, here and below we call $H$ to be the Hamiltonian, although it is actually the Hamiltonian density while the Hamiltonian is given by the integral $\int d^3 x\, H$.} $H=\F^{(\geq2)}$ and four constraints $\F^{(1)}+\F^{(\geq2)}=0$ and $\F_i^{(\geq1)}=0$, where $(n,n^i)$ are Lagrange multipliers. We see that on the constraint surface $\F^{(1)}+\F^{(\geq2)}=0$, the Hamiltonian simplifies, i.e it 
only contains the "linear terms"
\be\label{linh}
H=-\F^{(1)}.
\ee
Of course, this is an apparently false oversimplification since the constraint surface must also be sliced by gauge fixing conditions. Still, the slicing can be done in such a way that the Hamiltonian can be determined exactly. Indeed, in the $\vf=0$ gauge one can explicitly determine  $p_\vf$ from the Hamiltonian constraint $\F^{(1)}+\F^{(\geq2)}=0$ as this is simply a quadratic polynomial in $p_\vf$. In this way the system can be deparametrized; setting $\vf=0$ and determining $p_\vf$ in terms of other variables correspond to choosing a time gauge. As we will discuss in the next section, the momentum constraint $\F_i^{(\geq1)}=0$ can be dealt with perturbatively in $\cc_{ij}$ and $\s^{ij}$ while keeping $\z$ and $\pi$ dependence exact.

Although our main interest in this paper is cosmology, one can generalize the above construction to any solution in general relativity (or to other theories with first class constraints and Lagrange multipliers). One can denote the background values and the fluctuations generically as $Q_i=\bar{Q}_i+\d Q_i$, $P_i=\bar{P}_i+\d P_i$ and $N^\m=\bar{N}^\m+\d N^\m$, where $N^\m$ collectively denotes  the lapse and the shift (or Lagrange multipliers in the theory which do not have conjugate momenta). The  constraints can be expanded in the fluctuation fields as above $\Phi_\m=\Phi_\m^{(1)}+\Phi_\m^{(\geq 2)}$. Using these expansions in the action should yield 
 \be
 S=\int d^4 x\left[\d P_i \d \dot{Q}_i\ - \bar{N}^\m \Phi_\m^{(\geq2)}-\d N^\m  \left(\Phi_\m^{(1)}+\Phi_\m^{(\geq2)}\right)\right],
 \ee
where the linear terms cancel and $\Phi_\m^{(0)}=0$ since the background is assumed to be a solution. Then, the Hamiltonian for the fluctuations becomes  $H= \bar{N}^\m \Phi_\m^{(\geq2)}$ and the new constraints imposed by $\d N^\m$ field equations read $\Phi_\m^{(1)}+\Phi_\m^{(\geq2)}=0$. Again, the Hamiltonian on the constraint surface simplifies to 
\be\label{linhg}
H=-\bar{N}^\m\, \F_\m^{(1)}.
\ee
Note that for a generic background, the new Hamiltonian \eq{linhg} gets contributions from both the original Hamiltonian and the momentum constraints ($\Phi_0$ and $\Phi_i$), while in the cosmological case \eq{linh}, only $\F_0$ contributes to $H$ since $\bar{N}^i=0$.

As we will see below, the rest of the computation depends on how one solves the constraints and imposes the gauge fixing to identify a set of basic physical degrees of freedom (in our case, these would be $\z$, transverse-traceless $\cc_{ij}$ and their conjugate momenta) and to determine the rest of the variables in terms of this basic set. Usually this is done perturbatively as first performed in the original work \cite{mal}. Here, we would like to go beyond the perturbation theory, therefore we will try to solve the equation $\F^{(1)}+\F^{(\geq2)}=0$ exactly without assuming any order between the terms. Of course, among different ways of solving this equation we would like to fix $\F^{(1)}$ in terms of the main physical variables we identified since this determines the Hamiltonian \eq{linh}. One may see \eq{pvf} below to observe how this works out, where $p_\vf$, which is in $\F^{(1)}$, is determined in terms of other variables; corresponding to the equation $\F^{(1)}=-\F^{(\geq2)}$.

\section{Deparametrized semi-exact Hamiltonian}

We now carry out the above computation explicitly. In terms of the variables introduced in \eq{pert},  the Hamiltonian constraint becomes
\bea
\F=&&\fr{2e^{-3\z}}{a^3}\left[-\fr{1}{24}(\pi-6a^2\dot{a})^2+\s^{ij}\s^{kl}\hat{\cc}_{ik}\hat{\cc}_{jl}\right]+\fr{e^{-3\z}}{2a^3}\left[p_\vf+a^3\dot{\f}\right]^2\nn\\
&&+a^3e^{3\z}\left[V(\f(t)+\vf)+\fr{e^{-2\z}}{2a^2}\hat{\cc}^{ij}\del_i\vf\del_j\vf-\fr12 R^{(3)}\right].
\eea
The momentum $p_\vf$ can be solved exactly from $\F=0$ as 
\be\label{pvfi}
p_\vf=-a^3\dot{\f}\pm\left[\fr16(\pi-6a^2\dot{a})^2-4\s^{ij}\s^{kl}\hat{\cc}_{ik}\hat{\cc}_{jl}-2a^6e^{6\z}\left(V(\f(t)+\vf)+\fr{e^{-2\z}}{2a^2}\hat{\cc}^{ij}\del_i\vf\del_j\vf-\fr12 R^{(3)}\right)\right]^{1/2},
\ee
where $\pm$ corresponds to the two different roots of the quadratic equation. This ambiguity is precisely related  the so called notorious problem of time  that arises in the Hamiltonian formulation of general relativity. To solve it, we first assume $\dot{\f}<0$ in the classical solution, which is the case for the slow-roll inflation, and demand  $p_\vf$ vanishes when other perturbations are turned off, which implies
\be\label{pvf}
p_\vf=-a^3\dot{\f}-\left[\fr16(\pi-6a^2\dot{a})^2-4\s^{ij}\s^{kl}\hat{\cc}_{ik}\hat{\cc}_{jl}-2a^6e^{6\z}\left(V(\f(t)+\vf)+\fr{e^{-2\z}}{2a^2}\hat{\cc}^{ij}\del_i\vf\del_j\vf-\fr12 R^{(3)}\right)\right]^{1/2}.
\ee
Therefore, the  problem of time is solved by referring to the specific (inflationary) background.   Note that \eq{pvf} is valid as long as the square root is meaningful, which is the only restriction (that is automatically satisfied in perturbation theory).  We will always assume that \eq{pvf} is well defined although obviously one can imagine field configurations where the term in the square root becomes negative. In that case one should revise the solution \eq{pvf} and reconstruct the Hamiltonian below. 

As discussed in the previous section, one also needs the linearized constraint $\F^{(1)}$ that gives the Hamiltonian \eq{linh}. For that  one may note 
\be\label{r17}
R^{(3)}=\fr{e^{-2\z}}{a^2}\left[R^{(3)}(\hat{\cc})-4\hat{\cc}^{ij}\hat{\nabla}_i\hat{\nabla}_j\z-2\hat{\cc}^{ij}\del_i\z\del_j\z \right],
\ee
where $R^{(3)}(\hat{\cc})$ and $\hat{\nabla}$ are the Ricci scalar and the covariant derivative of $\hat{\cc}_{ij}$. A straightforward calculation then gives 
\be
\F^{(1)}=-\fr{1}{2}\,a\,\del_i\del_j\cc_{ij}+2\,a\,\del^2\z+6Va^3\z+a^3\,\fr{\del V}{\del\f}\,\vf+\dot{\f} p_\vf+\fr{\dot{a}}{a}\pi,\label{f1}\\
\ee
where $V$ and $\del V/\del \f$ are functions evaluated on the background field $\f(t)$, e.g. $V=V(\f(t))$. 

At this point one can impose the gauge 
\be
\vf=0, \label{gf1}
\ee
and use \eq{linh}, \eq{pvf} and \eq{f1} to obtain  
\bea
H=&&a^3\dot{\f}^2+\fr{1}{2}\,a\,\del_i\del_j\cc_{ij}-2\,a\,\del^2\z-6Va^3\z-\fr{\dot{a}}{a}\pi\nn\\
&&+\dot{\f} \left[a^6\dot{\f}^2+\fr16\pi^2-2a^2\dot{a}\pi-4\s^{ij}\s^{kl}\hat{\cc}_{ik}\hat{\cc}_{jl}-2a^6V(e^{6\z}-1)+a^6e^{6\z} R^{(3)}\right]^{1/2}.   \label{nph}
\eea
One may expand this Hamiltonian in the fluctuation fields to see that the zeroth and the first order terms cancel each other in $H$, as it should be.

It is important to emphasize that the gauge fixing \eq{gf1} breaks down when the scalar background  obeys $\dot{\f}=0$  (yet it is possible to eliminate $\vf$ and $p_\vf$ via other smooth gauges which are valid even when $\dot{\f}=0$, see \cite{ak1}).  In many models $\dot{\f}=0$ happens repeatedly while the scalar oscillates about the minimum of its potential during rehating. In that case the expansion of the square-root in \eq{nph} around $a^6\dot{\f}^2$ fails, which is directly related to the breakdown of the gauge fixing condition \eq{gf1}.  

To proceed one should work out  the momentum constraint. In terms of the fluctuation fields introduced in \eq{pert}, one can see that the momentum constraint \eq{c1} becomes 
\be\label{mcs}
\hat{\cc}_{ik}\hat{\nabla}_j\s^{kj}=-\fr16\del_i\pi+\fr12(-6a^2\dot{a}+\pi)\del_i\z,
\ee
where we have also utilized the gauge \eq{gf1} (recall that $\hat{\nabla}$ is the derivative operator of $\hat{\cc}_{ij}$).  We would like to underline that no approximation is done in obtaining \eq{mcs}. The momentum tensor $\s^{ij}$ can be decomposed as
\be\label{sdec}
\s^{ij}=\s^{ij}_{TT}+\hat{\nabla}^i\s^j_T+\hat{\nabla}^j\s^i_T+\hat{\nabla}^i\hat{\nabla}^j\s-\fr13\hat{\cc}^{ij}\hat{\nabla}^k\hat{\nabla}_k\s;
\ee
where $\hat{\nabla}_i\s^{ij}_{TT}=0$ and $\hat{\nabla}_i\s^{i}_{T}=0$. One can see that $\s^{ij}_{TT}$ drops out from the constraint equation \eq{mcs} and thus it becomes the physical momentum conjugate to $\cc_{ij}$.  The auxiliary fields $\s_T^i$ and $\s$ can (in principle) be determined from \eq{mcs} uniquely once appropriate boundary conditions are imposed. The evolution will be completely fixed once a corresponding gauge condition is imposed. We set
\be\label{gf2}
\del_i\cc_{ij}=0, 
\ee
which is the standard choice for the tensor field. 

The conditions \eq{gf1} and \eq{gf2} fix the coordinate gauge freedom completely. Since we are trying to make an exact computation, extra care is needed  when switching to a different gauge choice. In the linear theory, the gauge transformations related to coordinate changes are given by the Lie derivative of the background. In an exact treatment, this should be modified to include all higher order terms, which can be expressed as the exponentiation of the Lie derivative, see e.g. \cite{revmalik}.  

\subsection{Quadratic case} 

Before continuing, it is useful to make a cross check of the above computation at the quadratic order. As usual the tensor and the scalar fields decouple from each other and a straightforward  computation (which uses the quadratic expression for  $R^{(3)}(\hat{\cc})$ in the gauge \eq{gf2} and some integration by parts) gives the standard Hamiltonian for the tensor field 
\be
H^{(2)}_\cc= \frac{2}{a^3}\s^{ij}_{TT}\s^{ij}_{TT}+\fr{1}{8}a (\del_i\cc_{jk})(\del_i\cc_{jk}).
\ee
On the other hand, the momentum constraint \eq{mcs} yields
\be 
\s_T^i=0,\hs{5}\s=-\fr14\del^{-2}\pi-\fr{9a^2\dot{a}}{2}\del^{-2}\z,
\ee
which can be used in \eq{nph} to find the following quadratic Hamiltonian for the curvature perturbation 
\be\label{qhc}
H^{(2)}_\z=\fr{1}{2a^3\dot{\f}^2}\left[\fr{\dot{a}}{a}\pi+6Va^3\z+2a\del^2\z\right]^2+\fr{3\dot{a}}{a}\pi\z+\left[18Va^3+27 a\dot{a}\right]\z^2-7a\del_i\z\del_i\z.
\ee
Here,  $\del^{-2}$ stands for the Green function of the flat space Laplacian $\del^2$ and for a test function $f(\vec{x})$ it is given by 
\be\label{lgf}
\del^{-2}f(\vec{x})=-\fr{1}{4\pi}\int  \fr{d^3 y}{|\vec{x}-\vec{y}|} \, f(\vec{y}). 
\ee
One can see that $[\del^{-2},\del_i]=0$ provided the fields obey suitable boundary/fall-off conditions. Although \eq{qhc} does not look familiar, after the canonical transformation 
\be\label{ct0}
\pi\to\pi-18a^2\dot{a}\z-\fr{2a^2}{\dot{a}}\del^2\z
\ee
the Hamiltonian becomes 
\be\label{qhz}
H^{(2)}_\z=\fr12\fr{\dot{a}^2}{a^5\dot{\f}^2}\pi^2+ \fr12 \fr{\dot{\f}^2a^3}{\dot{a}^2}\del_i\z\del_i\z
\ee
which is again the standard expression. Note that \eq{ct0} is an explicitly time dependent canonical transformation and to find the Hamiltonian in the new variables one should make the substitution in the action $S=\int dt d^3x (\pi\dot{\z}-H)$ and perform some integration by parts. 

\subsection{Field redefinitions and shift symmetry}

The above quadratic computation reveals an important symmetry of the action,  the invariance under constant shifts of $\z$, 
\be
\z\to\z+c.
\ee
This symmetry is not explicit in \eq{nph}, as it is does not evident in the quadratic Hamiltonian \eq{qhz}, since $\pi$ must also be non-trivially transformed.  However, one can apply the canonical map 
\be\label{ct1} 
\pi\to \pi + 6a^2 \dot{a} - 6 a^2 \dot{a} e^{3\z}
\ee
to obtain
\be
H=a^3\dot{\f}^2e^{3\z}-\fr{\dot{a}}{a}\pi-\dot{\f}^2a^3 e^{3\z} \left[1+\fr{e^{-6\z}}{6\dot{\f}^2a^6}\pi^2-
\fr{2(\dot{a}/a)e^{-3\z}}{\dot{\f}^2a^3}\pi
-\fr{4e^{-6\z}}{\dot{\f}^2a^6}\s^{ij}\s^{kl}\hat{\cc}_{ik}\hat{\cc}_{jl}+\fr{1}{\dot{\f}^2}R^{(3)}\right]^{1/2}.   \label{nph2}
\ee
We see that here $\z$ and $a(t)$ only appear in the combination 
\be
a e^{\z},
\ee
(and $\dot{a}$ only appears in the Hubble parameter $\dot{a}/a$), hence shifting $\z$ by a constant (by keeping $\pi$ unchanged) corresponds to scaling the scale factor $a(t)$ by a constant, $a\to\l a$, which is a symmetry of the theory since this is equivalent to the coordinate change $x^i\to\l x^i$. 

After applying \eq{ct1}, the momentum constraint also simplifies a bit to take the form 
\be\label{mcs2}
\hat{\cc}_{ik}\hat{\nabla}_j\s^{kj}=-\fr16\del_i\pi+\fr12\pi\del_i\z.
\ee
We could not solve \eq{mcs2} for $\s^i_T$ and $\s$ in an exact and explicit way, and to our knowledge there is no known way of doing it. Yet, we see that $\z$ and $\pi$ appear only in the right hand side and the linear operator acting on $\s^{ij}$ only depends on $\cc_{ij}$. Therefore, \eq{mcs2} can be solved perturbatively in $\cc_{ij}$ by keeping $\z$ and $\pi$ dependence exact at each order. Substituting this solution back in \eq{nph2} gives the Hamiltonian for the physical degrees of freedom. 

It is possible to simplify the form of the exact Hamiltonian by a  canonical field redefinition
\be\label{upu}
u=a^3e^{3\z},\hs{7}P_u=\fr{e^{-3\z}}{3a^3}\pi,
\ee
which gives
\be
H=\dot{\f}^2 u-\dot{\f}^2 u \left[1+\fr{3}{2\dot{\f}^2}P_u^2-
\fr{6(\dot{a}/a)}{\dot{\f}^2}P_u
-\fr{4}{\dot{\f}^2 u^2}\s^{ij}\s^{kl}\hat{\cc}_{ik}\hat{\cc}_{jl}+\fr{1}{\dot{\f}^2}R^{(3)}\right]^{1/2}.   \label{nph3}
\ee
The momentum constraint \eq{mcs2} also simplifies after this change
\be\label{mcs3}
\hat{\cc}_{ik}\hat{\nabla}_j\s^{kj}=-\fr12u\, \del_iP_u. 
\ee
Note that the background time dependence is now encoded in the slowly varying factors $\dot{a}/a$ and $\dot{\f}$, therefore this form of the Hamiltonian can be useful in numerical computations or finding explicit classical solutions, see the section below. 

\subsection{Truncation to the scalar sector} 

If we focus on the truncation where the tensor modes are set to zero, $\cc_{ij}=0$ (i.e. $\hat{\cc}_{ij}=\d_{ij}$) and $\s^{ij}_{TT}=0$,  one has 
\be\label{sdec2}
\s^{ij}=\del_i \s^j_T+\del_j\s^i_T+\del_i \del_j \s-\fr13\d^{ij}\del^2\, \s
\ee
and \eq{mcs2} implies
\bea
&&\s_T^i=\fr12 \del^{-2}\left[ \pi\del_i\z\right]-\fr12\del_i \del^{-2} \del^{-2}\left[\del_j(\pi\del_j\z)\right] ,  \nn\\
&&\s=-\fr14 \del^{-2}\pi+\fr34 \del^{-2} \del^{-2}\del_j\left[\pi\del_j\z\right].\label{sst}
\eea
Using \eq{sst} and the expression for the Ricci scalar (corresponding to $\hat{\cc}_{ij}=\d_{ij}$)
\be
R^{(3)}=-\fr{e^{-2\z}}{a^2}\left[4\del^2\z+2\del_i\z\del_i\z \right]
\ee
in \eq{nph2} give the exact Hamiltonian for the canonical pair $(\z,\pi)$ when the tensor fields  are turned off. 

At this point, one may wonder if this truncation is consistent. This is an important question, which usually arises in Kaluza-Klein compactifications, where setting some  fields to zero is actually  inconsistent mathematically. What happens is that the field  equation of the  variable that is set to zero gets non-linear contributions from other fields and it is not trivially satisfied. In our case, one may check that setting  $\cc_{ij}=0$ and $\s^{ij}_{TT}=0$ is consistent since the corresponding equations does not receive any non-linear contributions from $\z$ and $\pi$, which is actually not possible due to the tensor structure. Moreover, the Poisson bracket algebra expressed in \eq{pb1} and \eq{pb2} also shows that the dynamics are separable (the truncation might be problematic if one would get a non-zero Poisson bracket say between $\z$ and $\s^{ij}$). 

The linear $-(\dot{a}/a)\pi$ term  in the exact Hamiltonian \eq{nph2} naively yields the equations of motion 
\be
\dot{\z}=-\fr{\dot{a}}{a}+...
\ee
and a logarithmic time dependence  $\z=-\ln(a)+...$ in the solution. This zeroth order term is actually canceled out in the perturbative expansion by a contribution coming from the square root, yet the higher order terms can be seen to have a similar structure. In a perturbative expansion in the canonical momenta, the leading order Hamiltonian becomes
\be
H=\fr{\dot{a}}{a}\left[\left(1+\fr{1}{\dot{\f}^2}R^{(3)}\right)^{-1/2}-1\right]\pi+...
\ee
Note that $\s^{ij}\s_{ij}$ is quadratic in $\pi$ or $\s^{ij}_{TT}$ and thus it is higher order in this expansion. The field equations now imply
\be\label{z44} 
\dot{\z}=\fr{\dot{a}}{a}\left[\left(1+\fr{1}{\dot{\f}^2}R^{(3)}\right)^{-1/2}-1\right]+...
\ee
where $R^{(3)}$ is given in \eq{r17}. In the minimal model considered here, this term is negligibly small since the curvature is suppressed by the Planck mass. However, if one adds, for instance, a (minimally coupled) spectator scalar field, its contribution to \eq{z44} is no longer suppressed by the Planck scale and this gives a meaningful $\ln(a)$ time dependence for $\z$ (see \cite{wisues} for an account on the IR logarithmic  loop corrections to the primordial scalar and tensor power spectra). The appearance of a logarithmic time dependence in the evolution of $\z$ is usually ascribed to IR loop effects. Nevertheless, it was shown in \cite{ak2} that in the minisuperspace quantum mechanics, where clearly no loop effects present, logarithmic terms also show up in the evolution of $\z$.  We see that there is a similar situation in the field theory case where the equations of motion directly imply an $\ln(a)$ dependence. Below we give some explicit solutions for $\z$ having this type of time evolution. Note that in the expansion of the square-root in the Hamiltonian \eq{nph2}, one encounters terms containing the powers of $(\dot{a}/a)\pi$. These  presumably yield $\ln(a)^n$ type of corrections to the power spectra discussed in the literature, see e.g. \cite{ir2}. 

\section{Long wavelength and late time limits} 

Obviously, it is difficult if not impossible to use \eq{nph2} in an exact way in the quantum theory. Yet \eq{nph2} is still useful since one can expand it to obtain all interactions at any given order. It is important to note that the quantum field theory of cosmological perturbations has issues, maybe the less severe one being the ordering ambiguity. The theory is also non-renormalizable (since it involves gravity) and there is no natural way of dealing with loop infinities; one simply subtracts, somehow arbitrarily, the large/infinite contributions to estimate the regularized loop corrections.  On the other hand, \eq{nph2} can be useful in determining the non-perturbative classical dynamics; especially the evolution of the long-wavelength superhorizon modes which are known to become classical to all orders \cite{cls}. However, there are subtleties in using \eq{nph2} because of the non-local expressions involving $\del^{-2}$ given in \eq{lgf}. In a naive derivative expansion of the action, the spatial Ricci scalar $R^{(3)}$ can be neglected but $\s^{ij}$ has derivative order zero,  see \eq{sdec2} and \eq{sst}, hence one should deal with the momentum constraint exactly if no other approximation is utilized. 

\subsection{Minisuperspace approximation} 

The first obvious way to proceed is to focus on the evolution of purely time dependent fields which can be specified by the ansatz $\z=\z(t)$, $\pi=\pi(t)$, $\cc_{ij}=\cc_{ij}(t)$ and $\s_{TT}^{ij}=\s_{TT}^{ij}(t)$ (where one still has the conditions $\det(\hat{\cc})=\det(\d_{ij}+\cc_{ij})=1$ and $(\d_{ij}+\cc_{ij})\s^{ij}_{TT}=0$); these can be thought to represent the zero modes. One may use this ansatz in the equations of motion, which is very similar to obtaining the background equations \eq{backsol} from the Einstein's field equations $G_{\m\n}=T_{\m\n}$; but one immediately sees that \eq{sst} becomes ill defined since $\del^{-2}$ diverges on the zero mode ansatz. Although, naively, the zero modes are defined as purely time dependent functions, they must actually be thought as zero momentum Fourier modes, e.g. $\z(t)=\lim_{q\to0}\z(t,\vec{q})$, where $\z(t,\vec{q})$ is the Fourier transform of $\z(t,\vec{x})$. The consequences of this limit in the linearized theory has been studied  in \cite{wad}.  Nevertheless, one can still consider a minisuperspace approximation where both the left and the hand right sides of \eq{mcs2} identically vanish when the fields are taken $x^i$ independent. In that case, one can also take $\s^i_T=0$ and $\s=0$ since there is no need to make the decomposition \eq{sdec} in the first place.  As a result, one can obtain the exact Hamiltonian 
\be
H=a^3\dot{\f}^2e^{3\z}-\fr{\dot{a}}{a}\pi-\dot{\f}^2a^3 e^{3\z} \left[1+\fr{e^{-6\z}}{6\dot{\f}^2a^6}\pi^2-
\fr{2(\dot{a}/a)e^{-3\z}}{\dot{\f}^2a^3}\pi
-\fr{4e^{-6\z}}{\dot{\f}^2a^6}\s_{TT}^{ij}\s^{kl}_{TT}\hat{\cc}_{ik}\hat{\cc}_{jl}\right]^{1/2}, \label{ms}
\ee
for the cosmological perturbations in the minisuperspace approximation. 

\subsection{Long wavelength boundary conditions} 

\begin{figure}
	\centerline{\includegraphics[width=9cm]{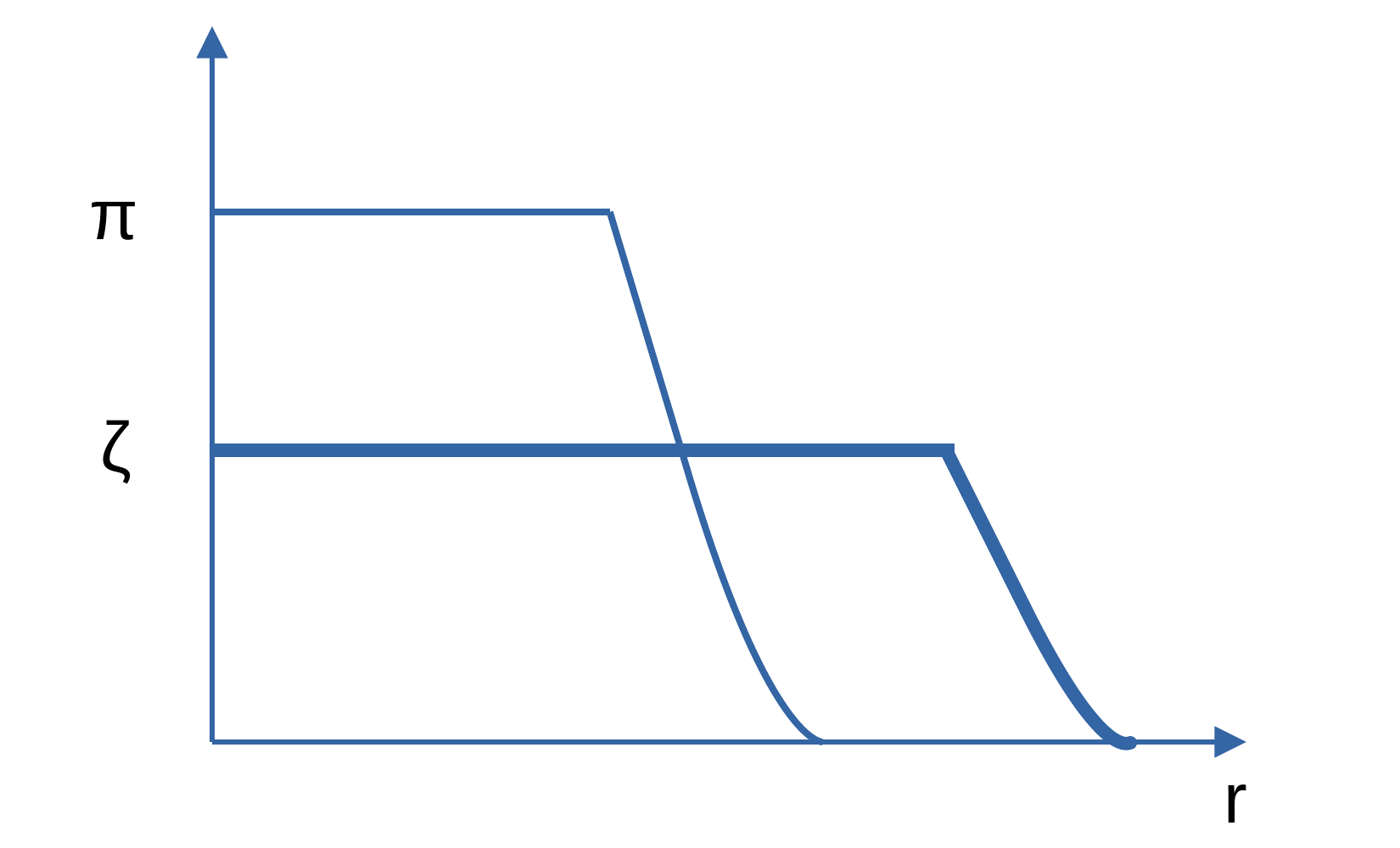}}
	\caption{A possible zero mode configuration for $\z$ and $\pi$. Note that $\pi$ vanishes in the region $\del_i\z\not=0$ thus one can set $\pi\del_i\z=0$. For such a configuration, the non-linear terms in \eq{sst} vanish giving $\s^i_T=0$ and $\s=-\del^{-2}\pi/4$.} 
	\label{fg1}
\end{figure}

The minisuperspace approximation can be used to have an overall understanding of the zero mode dynamics, but as we have mentioned above the actual physical configurations cannot  simply be time dependent fields because they must satisfy certain boundary conditions. In the linear theory, the issue hides in the Fourier transformation, which is usually taken to be granted but actually requires certain boundary conditions to be met. In the nonlinear theory, using the Fourier transformed momentum space modes does not help too much in a non-perturbative analysis due to mode mixing.  One way of dealing with the problem is to use the momentum constraint to glue different regions \cite{ne}. Here we will try to  define the zero modes as fields obeying appropriate boundary conditions. For example, one can imagine a zero mode field to be constant inside a ball of certain radius and drops smoothly but sharply to zero just outside (see Fig \ref{fg1}).  A singular version of this behavior can be written as  
\bea
&&\z(t,\vec{x})=\z(t)\,\th\left(\l_1-|\vec{x}|\right),\nn\\
&&\pi(t,\vec{x})=\pi(t)\,\th\left(\l_2-|\vec{x}|\right),
\eea
where $\th$ is the usual step function. Note that in the Hamiltonian analysis, one can take the coordinate and the momentum to obey different conditions since they represent independent degrees of freedom. 

If one takes\footnote{In the smooth version, $\l_1>\l_2$ corresponds to the case where  $\pi$ vanishes in the region $\z$ falls to zero from its constant value, see Fig \ref{fg1}.} $\l_1>\l_2$, then $\pi\del_i\z=0$ since $\pi=0$ when $\del_i\z\not=0$, and otherwise $\del_i\z=0$ already. In that case \eq{sst} implies $\s_T^i=0$ and $\s=-(1/4) \del^{-2}\pi$.  Then, one can see from \eq{sdec2} that $\s^{ij}=0$ when $|x|<\l_2$. Similarly, if one takes $\l_2>\l_1$, $\pi$ becomes spatially constant when $\del_i\z\not=0$, and thus for the function $\pi\del_i\z$ one can treat $\pi$ as if it is everywhere constant. In that case \eq{sst} gives again $\s_T^i=0$ and $\s=(1/2) \del^{-2}\pi$, and one finds  $\s^{ij}=0$ when $|x|<\l_1$. As we discussed above, setting $\s^{ij}=0$ (and taking the fields to be purely time dependent) gives the minisuperspace approximation and we therefore see that it is possible to imagine suitable boundary conditions so that the minisuperspace fields actually represent rigorous zero modes. 

\subsection{Late time limit} 

In an inflationary epoch where the scale factor increases nearly exponentially, the square-root of the exact Hamiltonian \eq{nph2} can be expanded in the inverse powers of $a$. Formally, the expansion works when
\be
ae^{\z}\gg1.
\ee
There are two different types of fall-off behavior in the square root in \eq{nph2}; the gradient terms coming from $R^{(3)}$ decrease like $1/a^2$ and the terms involving the momenta have the power $1/a^3$. By keeping track of these different types, the leading order late time Hamiltonian can be written as  
\bea
H=&&-2a\,e^{\z}\hat{\cc}^{ij}\del_i\z\del_j\z-a\,e^{\z}R^{(3)}(\hat{\cc})+{\cal O}\left(\fr{1}{a}\right)+
 \fr{Ve^{-3\z}}{6\dot{\f}^2a^3}\pi^2+\fr{2e^{-3\z}}{a^3}\s^{ij}\s^{kl}\hat{\cc}_{ik}\hat{\cc}_{jl}+{\cal O}\left(\fr{1}{a^6}\right)\nn\\
&&-\fr{\dot{a}/a\,e^{-2\z}}{2\dot{\f}^2a^2}\pi\,\left[R^{(3)}(\hat{\cc})-4\hat{\cc}^{ij}\hat{\nabla}_i\hat{\nabla}_j\z-2\hat{\cc}^{ij}\del_i\z\del_j\z \right] +{\cal O}\left(\fr{1}{a^4}\right)\label{loh} 
\eea
where we have used \eq{r17} and applied an integration by parts in the first line to get the first term (note that $\det \hat{\cc}_{ij}=1$).  This Hamiltonian captures the non-linear dynamics of the cosmological perturbations sometime after the beginning of inflation. 

The modes are supposed stretch over the horizon at late times and thus one may want to further apply a long wavelength approximation to \eq{loh}. Even in this simple case, the naive derivative expansion is nontrivial due to the non-local terms coming from $\s^{ij}$. To simplify the discussion we take the lowest order  solution \eq{sst} and further assume that the long wavelength modes obey the boundary conditions depicted in Fig \ref{fg1} so that $\pi\del_i\z=0$. After these simplifications, non-local terms related to  $\del^{-2}\pi$ still persist even after integration by parts. One way of proceeding is to discard the momentum constraint (by setting $\s^i_T=0$, $\s=0$ and taking all fields to be time dependent so that the momentum constraint is satisfied identically), which is the minisuperspace approximation that has been studied in \cite{ak2}. Here, we assume conditions like $\del_i\z\del_i\del_j\del^{-2}\pi\ll\del_j \pi$ which ensure the non-local terms that remain after integration by parts are negligible.  This yields the following late time long wavelength Hamiltonian
\be
H\to \fr{\dot{a}^2e^{-3\z}}{2\dot{\f}^2a^5}\pi^2+\fr{2e^{-3\z}}{a^3}\s_{TT}^{ij}\s^{kl}_{TT}\hat{\cc}_{ik}\hat{\cc}_{jl}, \label{loh2} 
\ee
which is actually the momentum part of the quadratic free Hamiltonian modified by the proper $e^\z$ factors. 

\section{Classical solutions}

The constant configuration $\z=\z_0$ and $\cc_{ij}=\cc_{ij}^0$ with zero momenta $\pi=0$ and $\s^{ij}_{TT}=0$ is an exact solution\footnote{This assertion assumes the momentum constraint is trivially satisfied with $\s^i_T=0$ and $\s=0$.} that supposedly describes the frozen out superhorizon modes. However this is an idealized description because this configuration is actually pure gauge and it can only be viewed as an asymptotic limit of a time dependent solution. It is possible to rigorously study the nonlinear evolution\footnote{One can view $\z(t)$ as a single isolated zero mode or as a collective field corresponding to a collection of long wavelength superhorizon modes, i.e. in the Fourier decomposition $\z(t,\vec{x})=\int_R d^3 k\, \exp(i\vec{k}.\vec{x})\,\z_k$  one can restrict a sufficiently small (comoving) IR region $R=(0,k_{IR})$ so that the approximation $\exp(i\vec{k}.\vec{x})\simeq1$ is a good one giving $\z(\t,\vec{x})\simeq\z(t)$.} of $\z(t)$ using the exact Hamiltonian.
 
\subsection{Late time evolution of the zero mode} 
 
We first use the late time long wavelength Hamiltonian \eq{loh2} and neglect the impact of the tensor mode on the curvature perturbation, which is expected to be suppressed by the Planck scale.  In that case one can exactly solve the equations that follow from \eq{loh2}  to obtain  
\bea
&&\z(t)=\z_i+\fr23 \ln\left[1+\pi_i \, e^{-3\z_i}\int_{t_i}^t \fr{3\dot{a}^2}{2a^5\dot{\f}^2}dt'\right],\nn\\
&&\pi(t)=\pi_i+\pi_i^2\,e^{-3\z_i}\int_{t_i}^t \fr{3\dot{a}^2}{2a^5\dot{\f}^2}dt'. \label{zsol}
\eea
This solution is very similar to the one obtained in the minisuperspace case \cite{ak2}. In an inflationary slow roll regime, one has $a\simeq a_i e^{Ht}$ and $\dot{\f}^2=2\e$, where $H$ and $\e$ are the (approximately constant) Hubble and the slow-roll parameters, respectively. For $t\gg t_i$ one finds to a very good approximation 
\bea
&&\z_f\simeq\z_i+\fr23 \ln\left[1+\fr{3H^2}{4\e} \, \fr{\pi_i}{a_i^3e^{3\z_i}}\right],\nn\\
&&\pi_f\simeq\pi_i+\fr{3H^2}{4\e} \, \fr{\pi_i^2}{a_i^3e^{3\z_i}}, \label{zsol2}
\eea
which shows that there is indeed a logarithmic change  of $\z$ that depends on the initial conditions $\z_i$ and $\pi_i$ imposed at some time $t_i$ presumably close to the beginning of inflation.  

\subsection{Some exact solutions in the minisuperspace model} 

The equations of motion can actually be integrated out explicitly in the minisuperspace approximation.  From the Hamiltonian given in \eq{nph3} one can find 
\bea
&&\fr{\dot{u}}{u}=-\fr{\fr32 P_u-3\dot{a}/a}{\left[1+\fr{3}{2\dot{\f}^2}P_u^2-\fr{6\dot{a}/a}{\dot{\f}^2}P_u\right]^{1/2}},\label{es1}\\
&&\dot{P}_u=-\dot{\f}^2+\dot{\f}^2\left[1+\fr{3}{2\dot{\f}^2}P_u^2-\fr{6\dot{a}/a}{\dot{\f}^2}P_u\right]^{1/2},\label{es2}
\eea
where one sets $\s^{ij}=0$ and $R^{(3)}=0$ for the minisuperspace approximation and $u$, $P_u$ are defined in \eq{upu}. We see that the equations  are decoupled, i.e. one can first integrate \eq{es2} to obtain $P_u$, which can then be used in \eq{es1} to determine $u$ as
\be
\ln u=-\int \fr{\fr32 P_u-3\dot{a}/a}{\left[1+\fr{3}{2\dot{\f}^2}P_u^2-\fr{6\dot{a}/a}{\dot{\f}^2}P_u\right]^{1/2}}dt.
\ee
Recall that $\z=\fr13 \ln u -\ln a$, see \eq{upu}. 

As emphasized above, these equations are only valid when the square root is well defined. This would require $P_u\leq P_u^-$ and $P_u\geq P_u^+$ where 
\be
P_u^\pm=2\fr{\dot{a}}{a}\pm2\sqrt{\fr{V}{3}}. 
\ee
Note that both the roots  $P_u^\pm$ are positive and $P_+<4\dot{a}/a$. 

Assuming $\dot{a}/a$ and $\dot{\f}$ can be treated as constants, which is a good approximation for an inflationary epoch, the solutions can be classified as follows:

If initially $0<P_u\leq P_-$, then \eq{es2} implies $\dot{P}_u<0$ always after that time. Therefore $P_u$ decreases in time until it eventually vanishes.   If initially $P_u<0$ then \eq{es2} implies $\dot{P}_u>0$ and $P_u$ increases to reach $P_u=0$ finally. Therefore for both of these cases $P_u=0$ is an attractor.  As $P_u\to0$, \eq{es1} shows that $u\to a^3$, which means $\z\to \z_0$, for some constant $\z_0$. This is the familiar case that also appears in perturbation theory where $\z$ eventually freezes out as in \eq{zsol}.  

If initially $P_+\leq  P_u <4\dot{a}/a$, then one has $\dot{P}_u<0$ thus $P_u$ flows through $P_+$. However at $P_u=P_u^+$, $\dot{P}_u=-\dot{\f}^2<0$, therefore this point cannot be an attractor. In this case, the evolution can no longer be described by the present equations and one should construct a new Hamiltonian by assuming the square root in \eq{pvf} is actually negative (i.e. one should redo the construction after taking a new solution for $p_\vf$ in \eq{pvf} so that the square root becomes meaningful).  

If initially $P_u=4\dot{a}/a$, then \eq{es2} implies $\dot{P}_u=0$, hence $P_u$ remains constant. Eq. \eq{es1} then shows $u=a^{-3}$ and $\z=-2\ln a$.

If, on the other hand,  one takes $P_u>4\dot{a}/a$ initially, \eq{es2} shows $\dot{P}_u>0$ afterwords and thus $P_u$ continuously increases.  In that case, \eq{es1} and \eq{es2} imply that  at large times
\be
u\to u_0\,e^{- \sqrt{\fr32}\,|\dot{\f}| \,t},\hs{5}P_u\to P_{u_0}\,e^{ \sqrt{\fr32}\,|\dot{\f}|\, t},
\ee
which gives asymptotically
\be
\z\to-\ln(a)-\sqrt{\fr16}\,|\dot{\f}| \,t.
\ee
We therefore see that the non-linearities inherited from general relativity give rise to highly nontrivial dynamics that naturally generates logarithmic $\ln(a)$ type time evolution as seen in the explicit solutions presented above. 

\section{Conclusions} 

In this paper, we reconsider the well-known expansion of cosmological perturbations around a FRW background in the Hamiltonian formalism and find out a non-perturbative way of dealing with the constraints and gauge fixing that leads to a non-perturbative semi-exact Hamiltonian involving the physical degrees of freedom. Our main observation is based on the fact that the full Hamiltonian in general relativity actually vanishes and a perturbative Hamiltonian appears from a first class constraint after the expansion around a background solution. By carefully analyzing this expansion, we show that on the constrained phase space of the physical degrees of freedom, the Hamiltonian only contains linear unconstrained fluctuations.  We then show that the theory can be deparameterized by solving the Hamiltonian constraint exactly where the problem of time is solved by referring to the background solution. This procedure finally yields a non-perturbative  Hamiltonian for the cosmological perturbations where the momentum constraint is solved perturbatively. We determine the the long wavelength and the late time asymptotics of the Hamiltonian and study the classical dynamics of the curvature perturbation zero mode.  

The present work can be useful in dealing with non-linearities in cosmological perturbation theory. Eq. \eq{nph2} can be expanded to obtain all $\z$ self interactions at any desired order which can then be used in the  in-in perturbation theory to calculate loop corrections to cosmological correlation functions. Note that in-in perturbation theory is naturally employed in the Hamiltonian formalism \cite{w1}. In the classical regime,  \eq{nph2} can be helpful in studying non-linear effects using analytical or numerical methods. 

As indicated earlier, it is difficult to use the exact Hamiltonian involving a square root of the fields in the quantum  theory without making any approximations (the usual expansion in the number of fluctuation fields is one such approximation). Moreover, as the theory inherently contains gravitational excitations, the renormalization  of the infinities is somehow ambiguous. Therefore, it would be  safer to assume that the exact Hamiltonian is valid in a semi-classical regime or in an effective field theory setting. Nevertheless,  one can try to utilize other approximations like finding non-trivial saddle points in a path integral quantization or one may  focus on  specific terms in the expansion (like the ones containing only the momentum $\pi$) and try to a re-sum the infinite series in the in-in perturbation theory. Additionally, quantization in the  minisuperspace approximation which yields a quantum mechanical system can be a good playground to deduce some exact results as discussed in \cite{ak2}.  In any case, some form of approximation is needed to use the exact Hamiltonian  in the quantum theory and consequently it looks difficult to verify the exact statements about the cosmological correlation functions that follow from various arguments like bootstrap by using the Hamiltonian obtained in this work. 

Finally, the formalism presented in this paper can easily be generalized to include  matter and (with some care about gauge fixing) to study the epoch of reheating since the analysis is actually valid on any FRW solution. Although we worked out the cosmological case here, the main idea can obviously be applied to different problems in general relativity that involve fluctuations around classical backgrounds. Among them, it would be interesting to see how the computation carries out for the Schwarzschild black hole, which we hope to address in a future work.

\end{document}